\documentclass[aip,jcp,twocolumn,reprint,superscriptaddress]{revtex4-2}

\usepackage{amsmath,amssymb,amsfonts}
\usepackage{bm}
\usepackage{graphicx}
\usepackage{physics}
\usepackage{braket}

\usepackage{xcolor}

\definecolor{erbblue}{RGB}{0,60,130}

\newcommand{\erbedit}[1]{#1}

\begin{document}




\title{Mixed response geometry and critical crossover in the Ising model}
\author{Eric R. Bittner}
\email{ebittner@central.uh.edu}
\affiliation{Department of Physics, University of Houston, Houston, Texas 77204, USA}
\affiliation{Institut Courtois \& D\'epartement de physique, Universit\'e de Montr\'eal, 1375 Avenue Th\'er\`ese-Lavoie-Roux, Montr\'eal H2V~0B3, Qu\'ebec, Canada}

\date{\today}

\begin{abstract}
We develop a geometric formulation of thermodynamic response in interacting spin systems and apply it to the two-dimensional Ising model. Treating inverse temperature and magnetic field as coordinates on a thermodynamic control manifold, we show that the mixed response field
$
\Omega_{\beta h}
=
-N\,\mathrm{cov}(m,e)
$
arises naturally as a curvature-like quantity that measures correlations between magnetic and energetic fluctuations. Monte Carlo simulations reveal a strongly localized mixed-response ridge that emerges from the critical point and extends into the finite-field crossover regime. Analysis of the susceptibility, specific heat, and mixed-response maxima demonstrates distinct scaling behavior in the magnetic, energetic, and mixed fluctuation sectors. When represented in normalized response coordinates, trajectories obtained at different magnetic fields collapse onto a common curve, indicating that the evolution of the mixed response is strongly constrained by the susceptibility. This collapse suggests the emergence of a low-dimensional response manifold and points toward a geometric description of critical crossover based on relations among response functions rather than equilibrium states alone. The framework establishes a direct connection between fluctuation correlations, critical scaling, and geometric thermodynamic response. 

\end{abstract}


\maketitle

\section{Introduction}\label{sec:I}

Many central relations of thermodynamics admit a natural geometric interpretation. Equilibrium thermodynamics may be formulated in terms of contact, symplectic, and information-geometric structures, where thermodynamic potentials, fluctuations, and stability properties define geometric objects on manifolds of equilibrium states.\cite{Weinhold1975,Ruppeiner1995,Grmela2014,vanDerSchaft2018} These approaches have provided valuable insight into phase transitions, critical phenomena, and the structure of thermodynamic state space.

Critical phenomena provide a particularly rich setting in which to explore such ideas. Near a critical point, thermodynamic response functions become strongly enhanced as fluctuations develop over increasingly large length scales. Quantities such as the magnetic susceptibility and heat capacity characterize the response of a system to external perturbations and provide the conventional signatures of critical behavior. From this perspective, criticality is fundamentally a problem of response.

Most geometric formulations of thermodynamics focus on the geometry of equilibrium states. The Weinhold and Ruppeiner metrics characterize fluctuations through derivatives of thermodynamic potentials, while information-geometric approaches employ measures of distinguishability and statistical distance to characterize state manifolds.\cite{Weinhold1975,Ruppeiner1995,ProvostVallee1980,Zanardi2007,Paris2009} In each case, geometry is associated primarily with the properties of thermodynamic states.

Here we adopt a complementary viewpoint in which geometry arises from thermodynamic response\cite{Bittner2026ab,Bittner2026_jcp}. Rather than constructing geometric objects from equilibrium states, we examine the relationships among response functions defined over the space of externally controllable parameters. This distinction shifts attention from the geometry of states to the geometry of fluctuations and response. Such a perspective is particularly natural near critical points, where multiple fluctuation channels become strongly coupled and response functions exhibit highly organized structure throughout parameter space.

A particularly useful quantity in this context is the mixed thermodynamic response
\begin{equation}
\Omega_{\beta h}
=
-\left(\frac{\partial m}{\partial\beta}\right)_h,
\label{eq:introOmega}
\end{equation}
which characterizes the sensitivity of the magnetization to changes in temperature at fixed field. Unlike the susceptibility and heat capacity, which probe fluctuations within individual thermodynamic sectors, $\Omega_{\beta h}$ couples magnetic and thermal response and therefore provides a direct measure of the interplay between energetic and magnetic degrees of freedom. As we show below, this mixed response develops highly nontrivial structure in the vicinity of a critical point.

The quantity $\Omega_{\beta h}$ also admits a natural geometric interpretation. Thermodynamic response functions are defined with respect to externally controllable parameters such as temperature, pressure, magnetic field, or chemical potential. These parameters span a control manifold on which response functions define local structure. In this representation, diagonal responses characterize fluctuations along individual directions in parameter space, while mixed responses quantify couplings between distinct directions. The resulting picture suggests that critical behavior may be organized not only through singular response functions, but also through geometric relationships among those responses.

Related ideas have emerged in several areas of statistical and quantum physics. Geometric descriptions based on fluctuation metrics, information geometry, and nonequilibrium response have proven useful for characterizing phase transitions, critical behavior, and dynamical evolution.\cite{BraunsteinCaves1994,Paris2009,Crooks1999,Seifert2012,Esposito2009} More recently, geometric structures associated with thermodynamic response have been explored in nonequilibrium systems and open quantum dynamics, where response functions naturally decompose into symmetric and antisymmetric sectors associated with fluctuations and path-dependent behavior. These developments motivate the broader question of whether critical response itself possesses an intrinsic geometric organization.

Critical phenomena are often characterized not only by singular behavior at a critical point, but also by the organization of response functions throughout the surrounding control manifold. Away from the critical point, thermodynamic response functions frequently develop maxima that define crossover trajectories extending into finite-field or supercritical regimes. Such structures have been discussed extensively in fluids, magnetic systems, and other many-body systems under the rubric of Widom lines and related crossover phenomena \cite{Xu2005WidomDynamic,Franzese2007WidomWater,Luo2014BehaviorWidom,Li2024SupercriticalCrossovers,Sordi2024QCDWidom}. While these constructions are traditionally associated with individual response functions, different fluctuation sectors need not generate identical crossover trajectories. Understanding how magnetic, thermal, and mixed fluctuations organize within the thermodynamic control manifold therefore remains an important open problem.

The geometric framework developed here provides a natural setting for addressing this question. By treating mixed response functions as fields defined on the control manifold, one may identify response ridges, trajectory relations, and scaling structures that emerge from the interplay of different fluctuation channels. As we show below, the mixed-response field generates crossover structures that are closely related to, but distinct from, those obtained from the susceptibility and heat capacity, revealing an underlying geometric organization of thermodynamic response.

To explore this question, we examine the two-dimensional Ising model as a benchmark system for studying mixed thermodynamic response. The Ising model provides an ideal setting because its critical behavior is well understood, allowing subtle features of the response structure to be isolated and analyzed quantitatively. We find that the mixed response field develops a strongly localized ridge that emerges from the critical point and extends into the finite-field crossover regime. This ridge identifies the region where magnetic and thermal response become most strongly coupled and defines a natural crossover trajectory in parameter space.

The mixed-response ridge is only the first indication of a deeper structure. By tracking the maxima of the susceptibility, specific heat, and mixed response, we construct a family of response ridges that characterize the magnetic, energetic, and mixed fluctuation sectors. These ridges exhibit distinct scaling behavior, revealing that different fluctuation channels organize differently as the critical point is approached. The resulting hierarchy provides a quantitative description of how thermal and magnetic fluctuations become coupled throughout the crossover region.

Beyond the scaling of individual response functions, we find that trajectories in normalized response space collapse onto an approximately universal curve over a broad range of applied fields. This collapse reveals a strong constraint linking the mixed response and susceptibility and demonstrates that the evolution of the response functions is far more structured than is apparent from the individual observables alone. Viewed geometrically, the collapse suggests that critical fluctuations evolve on a low-dimensional response manifold embedded within the higher-dimensional response space.

Taken together, these results support a response-based geometric description of critical phenomena that complements conventional state-space approaches. Rather than characterizing criticality solely through thermodynamic states and their fluctuations, the present framework emphasizes relationships among response functions themselves. The resulting structures provide new insight into how distinct fluctuation channels organize near criticality and suggest broader connections between thermodynamic response, scaling behavior, and geometry.

The remainder of this paper is organized as follows. Section \ref{sec:Framework} develops the geometric framework and introduces the mixed-response field. In Section ~\ref{sec:IsingModel} we 
apply this to a the Ising model 
and connect various responses to 
geometric quantities. 
Section~\ref{sec:MonteCarlo} describes the Monte Carlo methodology used to evaluate the mixed-response field and its crossover structure in the two-dimensional Ising model. Here we also examine the scaling behavior of the response ridges, the collapse of normalized response trajectories, and the resulting response-space organization. Finally, Section \ref{sec:Discussion} summarizes the principal results and discusses broader implications for geometric descriptions of critical phenomena.
\section{Geometric Formulation of Thermodynamic Response}
\label{sec:Framework}
\erbedit{
We formulate thermodynamic response on a control manifold
spanned by externally controllable parameters. This
framework provides a geometric interpretation of mixed
response functions and establishes the connection between
thermodynamic response, curvature, and geometric work.}

For a system described by an energy function $U(S,V)$, the First Law,
\begin{equation}
dU = T\,dS - P\,dV,
\end{equation}
defines a differential structure on thermodynamic state space. Local thermodynamic response is governed by mixed derivatives of the fundamental potential, such as
\begin{equation}
\Omega_{SV} = \frac{\partial^2 U}{\partial S \partial V},
\end{equation}
which characterize the coupling between thermodynamic variables and determine the response to infinitesimal variations.

More generally, for a system parameterized by a set of externally controlled variables $\lambda = (\lambda_1,\lambda_2)$, the work differential may be written as a one-form,
\begin{equation}
\delta W = A_i(\lambda)\, d\lambda_i,
\end{equation}
with associated curvature
\begin{equation}
\mathcal{F}_{ij} = \partial_{\lambda_i} A_j - \partial_{\lambda_j} A_i.
\end{equation}
The work performed over a closed cycle $C$ is then given by
\begin{equation}
W[C] = \oint_C A_i\, d\lambda_i = \iint_{\Sigma_C} \mathcal{F}_{ij}\, d\lambda_i \wedge d\lambda_j,
\end{equation}
identifying thermodynamic work as the flux of a curvature field defined over control-parameter space.

In this formulation, a nonzero work over a cycle arises precisely when the one-form $A_i(\lambda)\,d\lambda_i$ is non-exact, corresponding to a nonvanishing curvature $\mathcal{F}_{ij}$. The existence and structure of this curvature depend sensitively on the choice of control variables.

\paragraph{Geometric interpretation: work as holonomy.}

The work performed over a quasistatic cycle admits a natural
interpretation as a geometric holonomy. Writing
\begin{equation}
W[C]=\oint_C \mathcal A,
\end{equation}
where $\mathcal A=A_i(\lambda)d\lambda_i$ is a response
one-form, the associated curvature two-form
\begin{equation}
\Omega=d\mathcal A
\end{equation}
determines the work accumulated over the enclosed region,
\begin{equation}
W[C]
=
\iint_{\Sigma_C}\Omega .
\end{equation}
For infinitesimal cycles, the curvature may therefore be
identified as the work per unit area enclosed in control
space. In this sense, thermodynamic work constitutes a
geometric holonomy and the curvature provides a local
measure of thermodynamic response.

This relation provides an operational interpretation of the curvature as a local density of geometric work. In this sense, quasistatic thermodynamic work constitutes a geometric holonomy, with the curvature encoding the local structure of thermodynamic response. Viewed in this way, the curvature defines a response field over the control manifold, allowing thermodynamic response, fluctuation correlations, and crossover structure to be analyzed within a common geometric framework.

In the following sections, we apply this framework to the
two-dimensional Ising model and show that the mixed response
field generates a rich response structure that includes
localized crossover ridges, distinct scaling behavior, and
emergent organization in response space.


\section{Geometric Structure of the Ising Model}\label{sec:IsingModel}

We consider the two-dimensional Ising model with Hamiltonian
\begin{equation}
H(J,h) = -J\,\mathcal{B} - h\,\mathcal{M},
\end{equation}
where $\mathcal{B} = \sum_{\langle ij\rangle} s_i s_j$ and $\mathcal{M} = \sum_i s_i$.
The equilibrium state is given by the Gibbs distribution $\rho \propto e^{-\beta H}$,
with partition function
\begin{equation}
Z(\beta,h) = \sum_{\{s_i\}} e^{-\beta H},
\end{equation}
and free energy
\begin{equation}
F(\beta,h) = -\frac{1}{\beta}\ln Z(\beta,h).
\end{equation}

The thermodynamic response depends on how the system is parameterized in control space.
We therefore distinguish two natural manifolds: the coupling--field manifold $(J,h)$ at
fixed $\beta$, and the thermodynamic control manifold $(\beta,h)$ at fixed $J$.

\subsection{Flat and Curved Control Manifolds: $(J,h)$ vs $(\beta,h)$}

The emergence of a nontrivial curvature field depends sensitively on the choice of
control variables.

\paragraph{The $(J,h)$ manifold at fixed $\beta$.}
At fixed inverse temperature, the free energy defines an exact differential,
\begin{equation}
dF = -\langle \mathcal{B} \rangle\, dJ - \langle \mathcal{M} \rangle\, dh,
\end{equation}
so that the associated one-form
\begin{equation}
\mathcal{A}_{Jh} = -\langle \mathcal{B} \rangle\, dJ - \langle \mathcal{M} \rangle\, dh
\end{equation}
satisfies $\mathcal{A}_{Jh} = dF$. The curvature therefore vanishes identically,
\begin{equation}
\Omega_{Jh} = d\mathcal{A}_{Jh} = 0.
\end{equation}
Thus, the $(J,h)$ manifold is globally integrable: no geometric work can be generated.

\paragraph{The $(\beta,h)$ manifold at fixed $J$.}
When $\beta$ is treated as a control parameter, the work associated with magnetic cycles
is determined by the magnetization,
\begin{equation}
\mathcal{A} = -M(\beta,h)\, dh.
\end{equation}
The corresponding curvature two-form is
\begin{equation}
\Omega = d\mathcal{A} = \Omega_{\beta h}\, d\beta \wedge dh,
\end{equation}
with
\begin{equation}
\Omega_{\beta h}
= - \left(\frac{\partial M}{\partial \beta}\right)_h
= - \frac{\partial^2 F}{\partial \beta \partial h}.
\end{equation}

Using standard identities for canonical averages, the mixed derivative may be expressed
in terms of equilibrium fluctuations. For an observable $O$,
\begin{equation}
\frac{\partial}{\partial \beta} \langle O \rangle
= - \left( \langle O H \rangle - \langle O \rangle \langle H \rangle \right).
\end{equation}
Applying this identity to the magnetization $M = \langle \mathcal{M} \rangle$ yields
\begin{equation}
\Omega_{\beta h}
= \langle \mathcal{M} H \rangle - \langle \mathcal{M} \rangle \langle H \rangle.
\end{equation}
Introducing the intensive variables $m = \mathcal{M}/N$ and $e = H/N$, we obtain
\begin{equation}
\Omega_{\beta h}
= - N \left( \langle m e \rangle - \langle m \rangle \langle e \rangle \right).
\label{eq:curvature_cov}
\end{equation}

This result shows that the curvature is proportional to the covariance between energy
and magnetization fluctuations. The geometric response is therefore governed by
correlated fluctuations of the conjugate observables and is enhanced in regions where
these fluctuations become strongly coupled, such as near the critical point.

\subsection{Origin of Geometric Response}

The origin of geometric response may be understood directly from the differential
relations governing the thermodynamic control manifold. On the $(T,h)$ manifold, the
mixed response may be written as
\begin{equation}
\Omega_{Th}
= -\left(\frac{\partial M}{\partial T}\right)_h
= -\left(\frac{\partial S}{\partial h}\right)_T,
\end{equation}
where the second equality follows from the Maxwell relation implied by
$dF = -S\,dT - M\,dh$. Together with the heat capacity at fixed field,
\begin{equation}
C_h = T\left(\frac{\partial S}{\partial T}\right)_h,
\end{equation}
the entropy differential becomes
\begin{equation}
dS = \frac{C_h}{T}\,dT - \Omega_{Th}\,dh.
\label{eq:dS_response}
\end{equation}

Equation~(\ref{eq:dS_response}) shows that $C_h/T$ and $\Omega_{Th}$ are the two
components of the entropy gradient on the control manifold. The former measures the
thermal variation of the entropy, while the latter characterizes its field dependence.
The corresponding integrability condition,
\begin{equation}
\left(\frac{\partial C_h}{\partial h}\right)_T
= -T\left(\frac{\partial \Omega_{Th}}{\partial T}\right)_h,
\end{equation}
directly relates the field dependence of the heat capacity to the temperature dependence
of the mixed response.

Along an isentropic trajectory, $dS = 0$, Eq.~(\ref{eq:dS_response}) gives
\begin{equation}
d\ln T = \frac{\Omega_{Th}}{C_h}\,dh,
\label{eq:isentropic}
\end{equation}
showing that the ratio $\Omega_{Th}/C_h$ determines the slope of adiabatic curves in
control space.

The ratio $\Gamma_h \equiv \Omega_{Th}/C_h$ is recognized as the \emph{magnetic
Gr\"{u}neisen parameter}, which characterizes the magnetocaloric response of the system
and governs adiabatic temperature changes induced by variations in the applied
field.\cite{ZhuGarstRoschSi2003,GarstRosch2005} Near a critical point, $\Gamma_h$
diverges with an exponent set by the universality class, providing a sensitive
thermodynamic signature of criticality that is distinct from either the susceptibility
or the specific heat alone.\cite{ZhuGarstRoschSi2003,WuZhuSi2018} In the present
geometric framework, this divergence corresponds to the rapid growth of the mixed
response field $\Omega_{\beta h}$ relative to the heat capacity, and its locus of
extrema defines the mixed-response ridge examined below.

Likewise, the magnetization differential
\begin{equation}
dM = -\Omega_{Th}\,dT + \chi\,dh
\end{equation}
implies that constant-magnetization trajectories satisfy
\begin{equation}
\left(\frac{dh}{dT}\right)_M = \frac{\Omega_{Th}}{\chi}.
\label{eq:isomag}
\end{equation}

The ratio $\Omega_{Th}/\chi$ appearing in Eq.~(\ref{eq:isomag}) is likewise related to
the Gr\"{u}neisen parameter through the identity $\Omega_{Th}/\chi = \Gamma_h\,C_h/\chi$,
connecting the slope of isomagnetization contours to the relative divergence rates of
the susceptibility and heat capacity near criticality.\cite{GarstRosch2005}

These relations demonstrate that thermodynamic response functions do more than
characterize the magnitude of fluctuations. Through Eqs.~(\ref{eq:isentropic}) and
(\ref{eq:isomag}), they define local direction fields on the thermodynamic control
manifold and determine how thermodynamic observables evolve under changes in temperature
and field.

Near a critical point, the response functions become strongly enhanced and acquire
characteristic scaling behavior. As a consequence, the associated direction fields are
expected to organize into preferred crossover trajectories that emanate from the
critical region. The response ridges and scaling relations examined below may therefore
be viewed as manifestations of an underlying response geometry arising from the
interplay of magnetic, thermal, and mixed fluctuations.

\paragraph{Response as a Geometric Feature.}

The geometric structure described above arises directly from the organization of
thermodynamic response functions on the control manifold. The susceptibility, heat
capacity, and mixed response define local direction fields through
Eqs.~(\ref{eq:isentropic}) and (\ref{eq:isomag}), while their extrema identify regions
where fluctuations become most strongly enhanced. As a consequence, the geometry of the
control manifold is encoded not only in the local response coefficients but also in the
trajectories and crossover structures generated by them.

Near a critical point, thermodynamic response functions typically develop pronounced
maxima that extend into the finite-field regime. The loci of these maxima define
crossover trajectories that organize the surrounding response landscape and provide a
continuation of critical behavior away from the singular point itself. Different
response functions generally produce distinct trajectories, reflecting the fact that
magnetic, thermal, and mixed fluctuations need not scale identically.

From this perspective, response ridges represent geometric features of the thermodynamic
control manifold. Their locations, scaling behavior, and mutual relationships provide
information about how different fluctuation sectors interact and evolve near criticality.
In the numerical sections below, we examine these structures for the susceptibility,
heat capacity, and mixed response field and show that they reveal a highly organized
response geometry in the vicinity of the Ising critical point.

\begin{figure*}[t]
\centering
\includegraphics[width=0.85\textwidth]{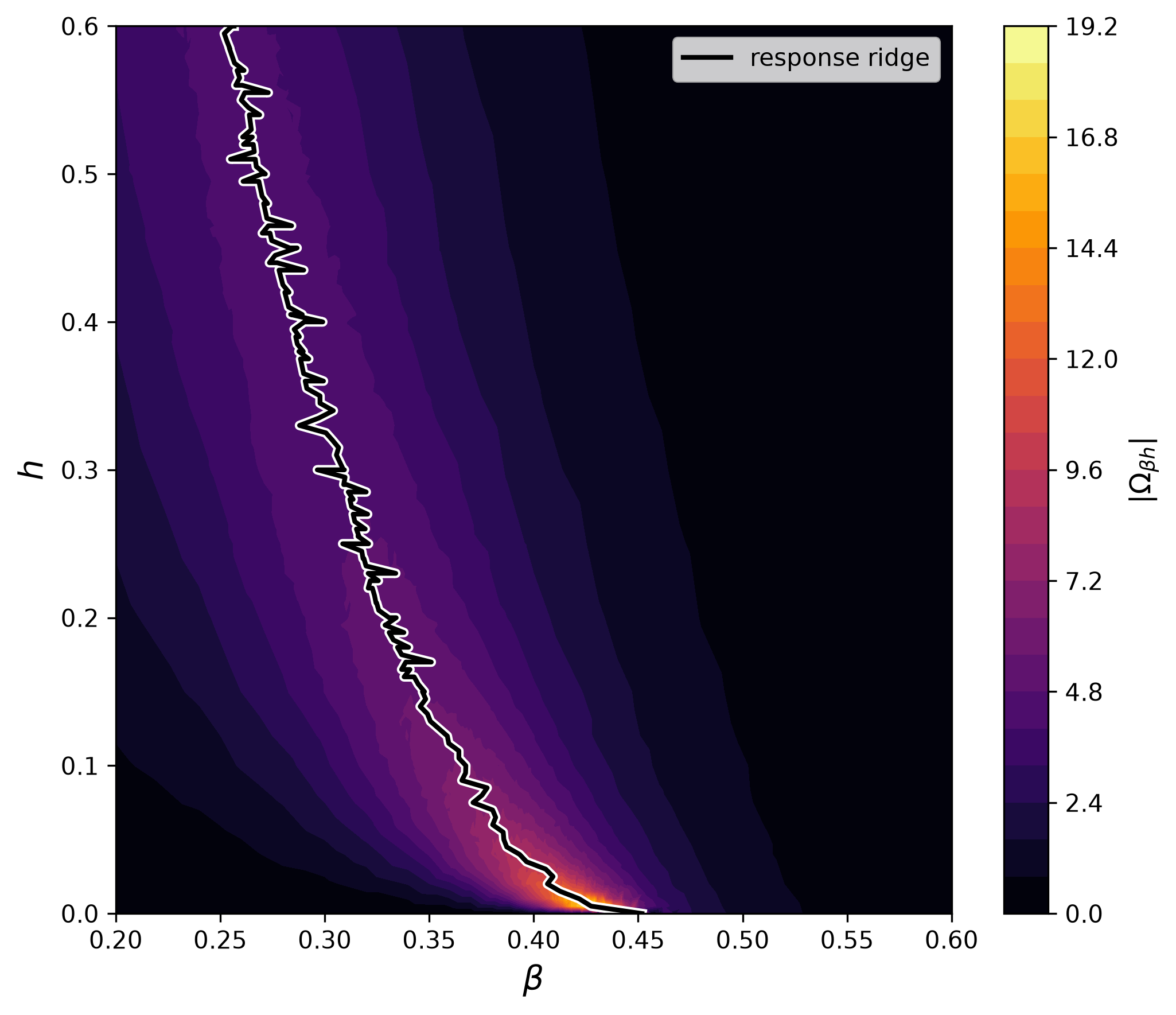}
\caption{
Magnitude of the mixed response field
$|\Omega_{\beta h}|=N|\mathrm{cov}(m,e)|$
for the two-dimensional Ising model as a function of inverse
temperature $\beta$ and magnetic field $h$.
The solid curve denotes the mixed-response ridge obtained
from the maxima of $|\Omega_{\beta h}|$ at fixed field.
The ridge originates at the critical point and extends into
the finite-field regime, defining a Widom-like trajectory
along which correlations between magnetic and energetic
fluctuations are strongest.
}
\label{fig:ising_curvature}
\end{figure*}

\section{Mixed response field and Critical Crossover}
\label{sec:MonteCarlo}

To evaluate the curvature field $\Omega_{\beta h}$ for the Ising model, we employ equilibrium Monte Carlo sampling of the canonical ensemble. As shown in the previous section, the curvature component may be expressed as
\begin{equation}
\Omega_{\beta h}
=
- N \left( \langle m e \rangle - \langle m \rangle \langle e \rangle \right),
\label{eq:curvature_estimator}
\end{equation}
which provides a direct estimator in terms of equilibrium fluctuations of the intensive magnetization $m$ and energy density $e$.

\subsection{Simulation protocol}

We consider the two-dimensional square-lattice Ising model with periodic boundary conditions and linear system size $L=64$, such that $N=L^2=4096$. For each point $(\beta,h)$ in control space, equilibrium configurations are generated using a standard Metropolis algorithm. After an initial thermalization period, expectation values $\langle m \rangle$, $\langle e \rangle$, and $\langle m e \rangle$ are computed via time averaging over Monte Carlo sweeps.

To resolve the structure of the curvature field, sampling is performed over a nonuniform grid in $(\beta,h)$, with increased density in regions where the curvature varies rapidly. In particular, additional sampling points are concentrated along the \erbedit{response}  ridge, as identified iteratively from preliminary simulations.

\erbedit{
Figure~\ref{fig:ising_curvature} shows the mixed response
field $|\Omega_{\beta h}|$ obtained from Monte Carlo
sampling. The response is strongly localized near the
critical point and forms a well-defined ridge that extends
into the finite-field regime. This ridge identifies the
region where magnetic and energetic fluctuations become most
strongly correlated.
}

\erbedit{
The ridge of maxima in $|\Omega_{\beta h}|$ defines a
field-dependent crossover trajectory that emanates from the
critical point. Similar trajectories arise from the maxima
of other response functions and provide a natural extension
of critical behavior into the finite-field regime.
}

\erbedit{
This construction is closely analogous to the Widom lines
used to characterize supercritical fluids and magnetic
systems, where the loci of response maxima serve as
extensions of the critical point into regions where true
thermodynamic singularities are absent. In the present case,
however, the ridge is defined by the mixed response
$|\Omega_{\beta h}|=N|\mathrm{cov}(m,e)|$ and therefore
tracks the crossover region where magnetic and energetic
fluctuations become most strongly correlated.
}

\erbedit{
The localization of $|\Omega_{\beta h}|$ demonstrates that
correlations between magnetization and energy are highly
structured within the control manifold. Rather than being
distributed uniformly throughout the $(\beta,h)$ plane, the
mixed response concentrates along a narrow crossover region
where magnetic and energetic fluctuations become most
strongly coupled. The resulting ridge identifies a preferred
path through parameter space and serves as the organizing
structure for the scaling and response-geometry analyses
presented below.
}

\erbedit{
While Fig.~\ref{fig:ising_curvature} reveals the existence of
a localized mixed-response ridge, it does not by itself
characterize the thermodynamic structure associated with that
ridge. To address this question, we examine the response
functions, scaling behavior, and trajectory structure of the
Ising model in greater detail below.
}

\subsection{Mixed Response and Curvature Ridges}

\begin{figure*}[t]
\centering
\includegraphics[width=\textwidth]{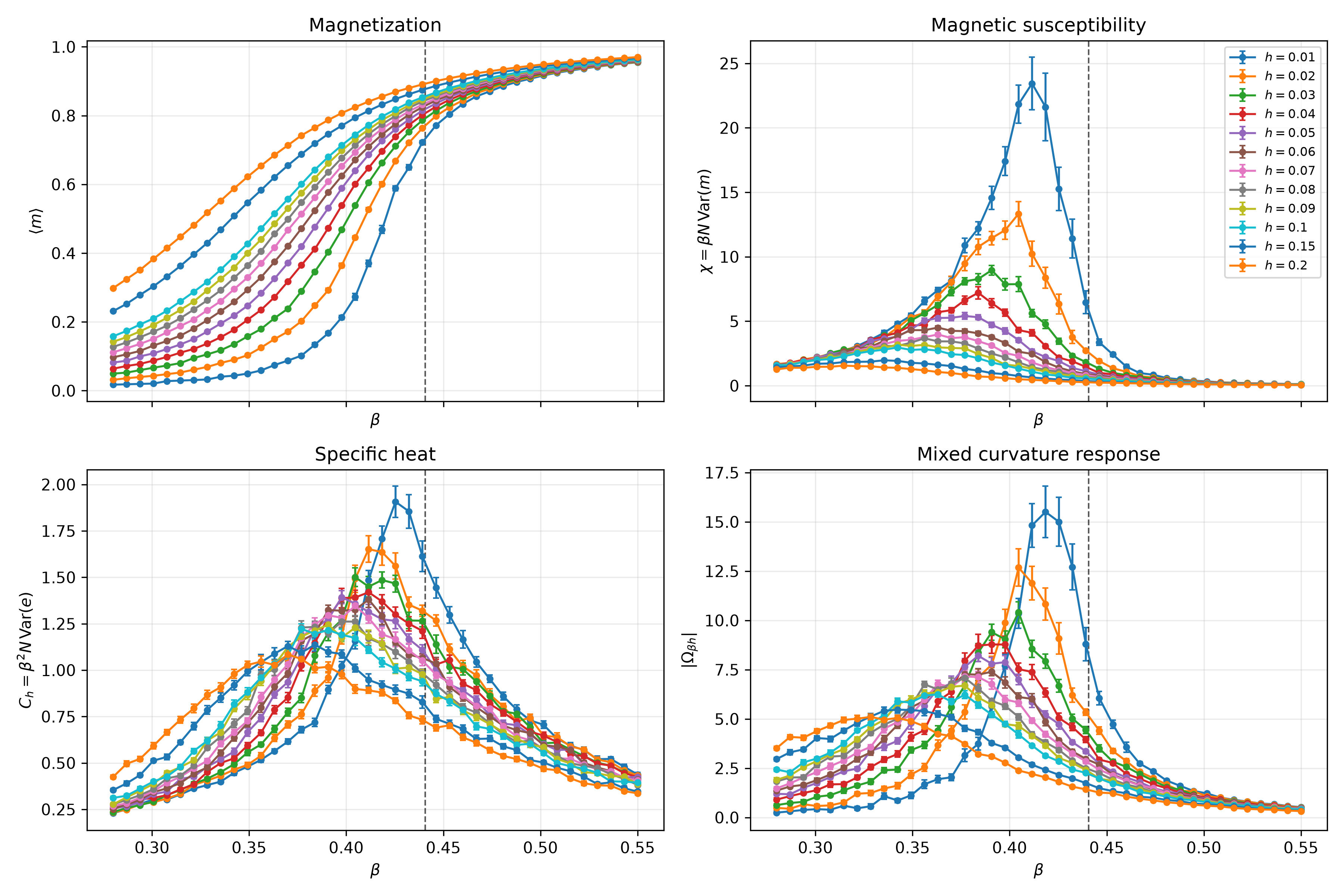}
\caption{
\erbedit{
Thermodynamic response functions of the two-dimensional Ising model as functions of inverse temperature $\beta$ for several applied magnetic fields $h$. Shown are the magnetization $\langle m\rangle$ (upper left), magnetic susceptibility $\chi=\beta N,\mathrm{Var}(m)$ (upper right), specific heat $C_h=\beta^2 N,\mathrm{Var}(e)$ (lower left), and the magnitude of the mixed response field $|\Omega_{\beta h}|$ (lower right). The vertical dashed line denotes the zero-field critical inverse temperature $\beta_c$. The field-dependent maxima define pseudocritical trajectories whose evolution is analyzed in the following sections.
}
}
\label{fig:ising_cross_sections}
\end{figure*}

\erbedit{
Figure~\ref{fig:ising_cross_sections} summarizes the principal
response functions of the two-dimensional Ising model as
functions of inverse temperature for several applied magnetic
fields. Shown are the magnetization $\langle m\rangle$, the
magnetic susceptibility $\chi$, the specific heat $C_h$, and
the magnitude of the mixed response field
$|\Omega_{\beta h}|$.
}

\erbedit{
As the magnetic field decreases, the crossover between the
disordered and ordered regimes sharpens and approaches the
critical behavior of the zero-field system. The
susceptibility and specific heat exhibit pronounced maxima
that identify the temperature range over which magnetic and
energetic fluctuations are strongest.
}
\erbedit{
The mixed response field displays a similar enhancement but
captures distinct physics. Since 
$
|\Omega_{\beta h}|
$
 directly measures correlations between magnetic and
energetic fluctuations, its magnitude
identifies the
region in which these fluctuation sectors become most
strongly coupled.
}

\erbedit{
Several features are immediately apparent. The maxima of
$\chi$, $C_h$, and $|\Omega_{\beta h}|$ all move away from
$\beta_c$ with increasing field, but they do not follow
identical trajectories. The mixed response peak tracks the
susceptibility more closely than the specific heat,
indicating that magnetic fluctuations dominate the
cross-correlation structure. At the same time,
$|\Omega_{\beta h}|$ increases rapidly as the field
decreases, demonstrating that correlations between energy
and magnetization become increasingly important upon
approaching criticality.
}


\subsection{Scaling of Peak Response Functions}

\erbedit{
The response maxima identified in
Fig.~\ref{fig:ising_cross_sections} define a set of
field-dependent pseudocritical points that extend the
critical point into the finite-field regime. For each value
of the applied field, we extract the locations and heights
of the maxima in the susceptibility, specific heat, and
mixed response field. These quantities provide a direct
measure of how the different fluctuation sectors evolve as
the system approaches criticality.
}

\erbedit{
The emergence of field-dependent response maxima may be understood from the scaling structure of the Ising critical point. In mean-field theory, the critical point extends into the finite-field regime through a family of crossover trajectories that connect continuously to the spinodal lines of the ordered phase. Along these trajectories, response functions attain finite maxima whose amplitudes and locations obey characteristic scaling laws. The finite-field maxima therefore provide a practical means of probing critical behavior even when the true singularity at $h=0$ is inaccessible.
}

\erbedit{
For the Curie–Weiss model, the susceptibility ridge follows directly from the equation of state and is given by
}
\begin{equation}
m=\pm\sqrt{1-\frac{T}{T_c}},
\end{equation}
\erbedit{
which implies the scaling relation
}
\begin{equation}
h \propto |\beta_c-\beta^\ast|^{3/2}.
\label{eq:CW_ridge_scaling}
\end{equation}

\erbedit{
This result provides a useful reference for interpreting the numerical Ising data. Although the two-dimensional Ising model belongs to a different universality class, one expects the finite-field response ridges to approach analogous power-law forms as the critical point is approached. The primary question is therefore not whether scaling occurs, but whether the magnetic, energetic, and mixed response sectors exhibit distinct scaling behavior and what this reveals about the geometric organization of critical fluctuations.
}

\begin{figure}[t]
\centering
\includegraphics[width=\columnwidth]
{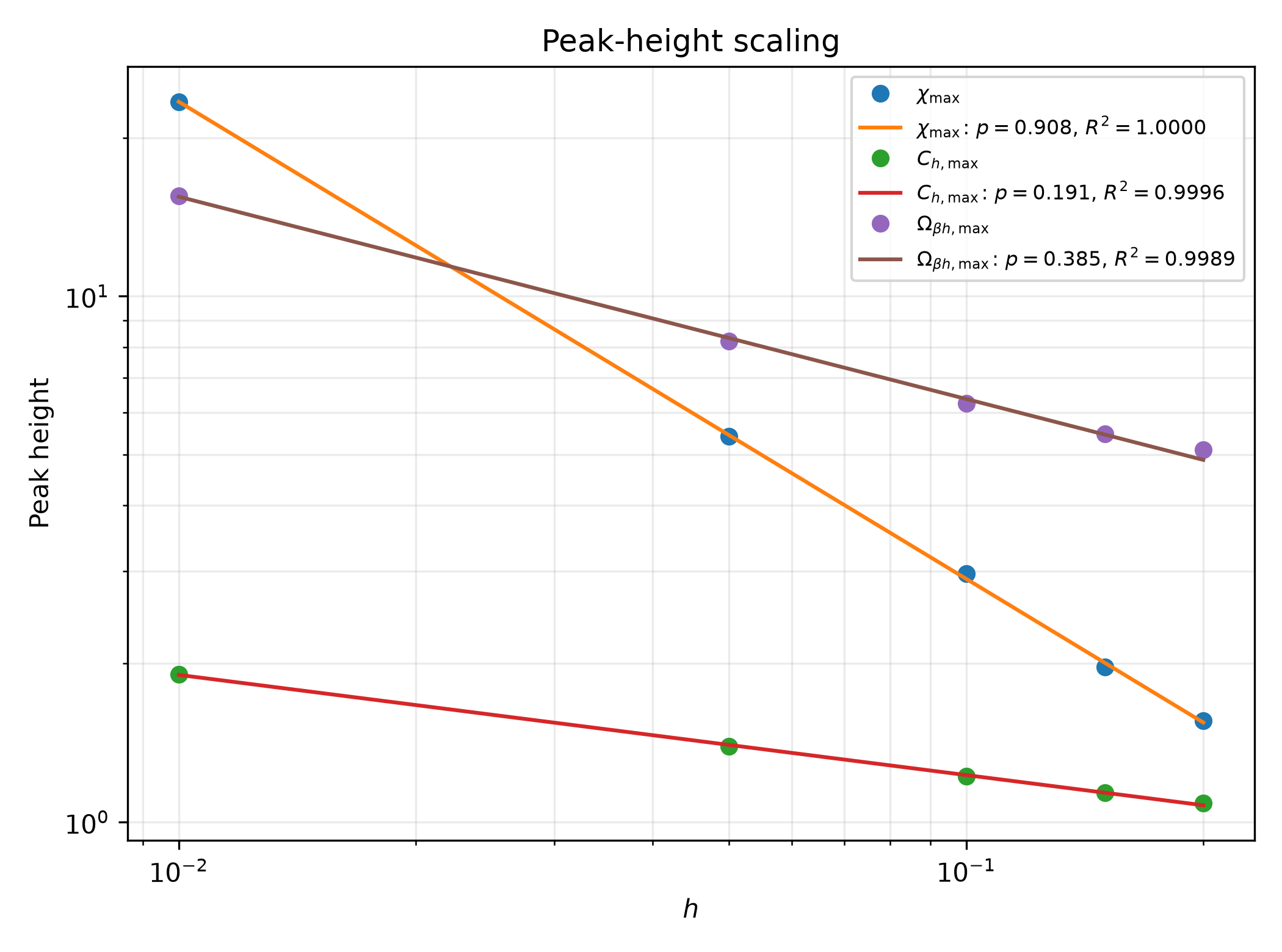}
\caption{
\erbedit{
Scaling of the peak response amplitudes with magnetic field.
The maxima of the susceptibility $\chi_{\max}$, specific heat
$C_{h,\max}$, and mixed response field
$|\Omega_{\beta h}|_{\max}$ are fit to
$Y_{\max}=Ah^{-p}$.
Solid lines denote least-squares power-law fits.
}
}
\label{fig:peak_height_scaling}
\end{figure}

\erbedit{
Figure~\ref{fig:peak_height_scaling} shows the dependence
of the peak response amplitudes on magnetic field. Over the
range of fields studied, all three observables are well
described by power-law scaling,
}
\begin{equation}
Y_{\rm max}(h) = A h^{-p},
\label{eq:peak_scaling}
\end{equation}
\erbedit{
where $Y_{\rm max}$ denotes the maximum value of the
corresponding response function.
}

\erbedit{
The susceptibility exhibits the strongest divergence,
with an exponent close to unity,
}
\[
\chi_{\max}\sim h^{-0.89}.
\]
\erbedit{
In contrast, the specific heat displays only a weak field
dependence,
}
\[
C_{h,\max}\sim h^{-0.19}.
\]
\erbedit{
The mixed response field lies between these two limits,
}
\[
|\Omega_{\beta h}|_{\max}\sim h^{-0.39},
\]
\erbedit{
demonstrating that the covariance between magnetic and
energetic fluctuations grows significantly as the critical
point is approached, although less rapidly than the
susceptibility itself.
}

\begin{table}[t]
\centering
\caption{
\erbedit{
Power-law fits for the peak response amplitudes and
pseudocritical ridge locations. Peak heights are fit to
$Y_{\max}=Ah^{-p}$, while ridge locations are fit to
$h=A|\beta_c-\beta^\ast|^p$. Uncertainties denote
one-standard-deviation fitting errors.
}
}
  
\label{tab:ising_scaling}
\begin{tabular}{lccc}
\hline
Observable & $A$ & $p$ & $R^2$ \\
\hline
\multicolumn{4}{c}{Peak heights: $Y_{\max}=Ah^{-p}$} \\
\hline
$\chi_{\max}$ &
$0.388 \pm 0.020$ &
$0.893 \pm 0.013$ &
0.9983 \\

$C_{h,\max}$ &
$0.797 \pm 0.015$ &
$0.187 \pm 0.006$ &
0.9902 \\

$|\Omega_{\beta h}|_{\max}$ &
$2.563 \pm 0.107$ &
$0.394 \pm 0.012$ &
0.9904 \\
\hline
\multicolumn{4}{c}{Ridge locations:
$h=A|\beta_c-\beta^\ast|^p$} \\
\hline
$\beta_\chi^\ast$ &
$19.32 \pm 7.93$ &
$2.22 \pm 0.18$ &
0.9613 \\

$\beta_C^\ast$ &
$24.27 \pm 22.66$ &
$1.85 \pm 0.33$ &
0.8241 \\

$\beta_\Omega^\ast$ &
$8.69 \pm 2.06$ &
$1.73 \pm 0.10$ &
0.9772 \\
\hline
\end{tabular}
\end{table}

\erbedit{
The locations of the response maxima also exhibit systematic
field dependence. Defining
$\beta_\chi^\ast$,
$\beta_C^\ast$,
and
$\beta_\Omega^\ast$
as the inverse temperatures at which the corresponding
response functions attain their maxima, we find that the
resulting pseudocritical trajectories are likewise
well described by power laws of the form
}
\begin{equation}
h = A|\beta_c-\beta^\ast|^p.
\label{eq:ridge_scaling}
\end{equation}

\erbedit{
The extracted exponents are summarized in
Table~\ref{tab:ising_scaling}. In all cases the response
ridges converge toward the zero-field critical point as
$h\rightarrow0$, confirming that the finite-field maxima
represent smooth continuations of the critical singularity.
}

The susceptibility ridge exponent $p_\chi = 2.22 \pm 0.18$ may be compared with the
exact theoretical prediction for the two-dimensional Ising universality class. From
scaling theory, the locus of susceptibility maxima in the $(h, T)$ plane obeys the
power law $h \propto |T - T_c|^{\Delta}$ as $h \to 0$, where $\Delta = \beta + \gamma$
is the gap exponent.\cite{AharonyFisher1983} For the two-dimensional Ising model, the
exact values $\beta = 1/8$ and $\gamma = 7/4$ give $\Delta = 15/8 = 1.875$, a result
that has been confirmed both by field-theoretic calculations and by high-precision
numerical studies.\cite{Mangazeev2010,LiJin2024} 
The fitted value $p_\chi = 2.22 \pm 0.18$ lies above this asymptotic prediction.
This discrepancy likely reflects a combination of finite-field and finite-size
corrections: the pseudocritical ridge has not yet converged to its asymptotic critical
scaling over the field range $h \in [0.01, 0.20]$ sampled here, and at $L = 64$ the
finite linear system size introduces corrections that shift the effective exponent
away from its thermodynamic-limit value.\cite{LiJin2024,Mangazeev2010} The exponent
$p_\chi$ is therefore expected to approach $\Delta = 15/8$ as $h \to 0$ and
$L \to \infty$, consistent with the two-dimensional Ising universality class.

\erbedit{
The ridge exponents reveal an interesting distinction between
the amplitudes and locations of the response maxima. The peak
heights of $|\Omega_{\beta h}|$ scale more similarly to the
susceptibility than to the specific heat, indicating that the
strength of the mixed response is primarily controlled by
magnetic fluctuations. In contrast, the ridge location
exponent $p_\Omega=1.73\pm0.10$ lies much closer to the
specific-heat exponent than to the susceptibility exponent.
This result suggests that while magnetic fluctuations
dominate the magnitude of the mixed response, energetic
fluctuations play a larger role in determining where the
cross-correlation reaches its maximum.
}

The mixed-response ridge occupies an intermediate position
between the magnetic and energetic sectors. Its scaling
exponent differs from those associated with either the
susceptibility or the specific heat, indicating that the
covariance field does not simply inherit the critical
behavior of one response function alone. Instead,
$\Omega_{\beta h}$ defines a distinct fluctuation sector
that emerges from the coupling between magnetic and
energetic degrees of freedom. The resulting ridge therefore
encodes information about the organization of correlated
fluctuations that is not accessible from either response
function individually.

Interestingly, the ridge exponent
$p_\Omega = 1.73 \pm 0.10$
lies close to the Curie--Weiss prediction of $3/2$.
Although the two-dimensional Ising model belongs to a
different universality class, this result suggests that the
geometry of the finite-field crossover retains important
mean-field characteristics. 
The mixed-response ridge exhibits an effective scaling trajectory
that is numerically close to the Curie--Weiss crossover form.

The exponents reported here should  be understood
as effective finite-field exponents over the range of fields
and system sizes studied, rather than as asymptotic critical
exponents. A full finite-size scaling analysis would be
required to determine the thermodynamic-limit values of the
ridge exponents. For the present purpose, however, the key
observation is not the precise numerical value of any single
exponent, but the existence of a systematic hierarchy among
the magnetic, energetic, and mixed-response ridges.

\subsection{Geometry of Normalized Response Trajectories}

\erbedit{
The scaling analysis above characterizes how the magnitudes and locations of the response maxima evolve with field. Equally important, however, is the manner in which the response functions evolve relative to one another. To examine this relationship, we construct trajectories in response space by eliminating the temperature parameter and plotting the mixed response field against the susceptibility as the system evolves through the crossover region.
}

\erbedit{
The close relationship between the susceptibility and mixed
response field may be understood by considering contours of
constant magnetization in the $(\beta,h)$ plane. Since
}
\[
dm
=
\Omega_{\beta h}\, d\beta
+
\chi\, dh,
\]
\erbedit{
the slope of an isomagnetization trajectory is
}
\[
\left(\frac{dh}{d\beta}\right)_m
=
-\frac{\Omega_{\beta h}}{\chi}.
\]
\erbedit{
The ratio $\Omega_{\beta h}/\chi$ therefore quantifies the
relative response of the order parameter to temperature and
field perturbations and defines a local geometric direction
in control-parameter space.
}

\erbedit{
For each magnetic field, the trajectory is generated by varying the inverse temperature $\beta$ and recording the corresponding values of $\chi$ and $|\Omega_{\beta h}|$. In this representation, the thermodynamic evolution of the system is represented as a path in response space rather than in the conventional $(\beta,h)$ control-parameter plane.
}

\begin{figure}[t]
\centering
\includegraphics[width=\columnwidth]
{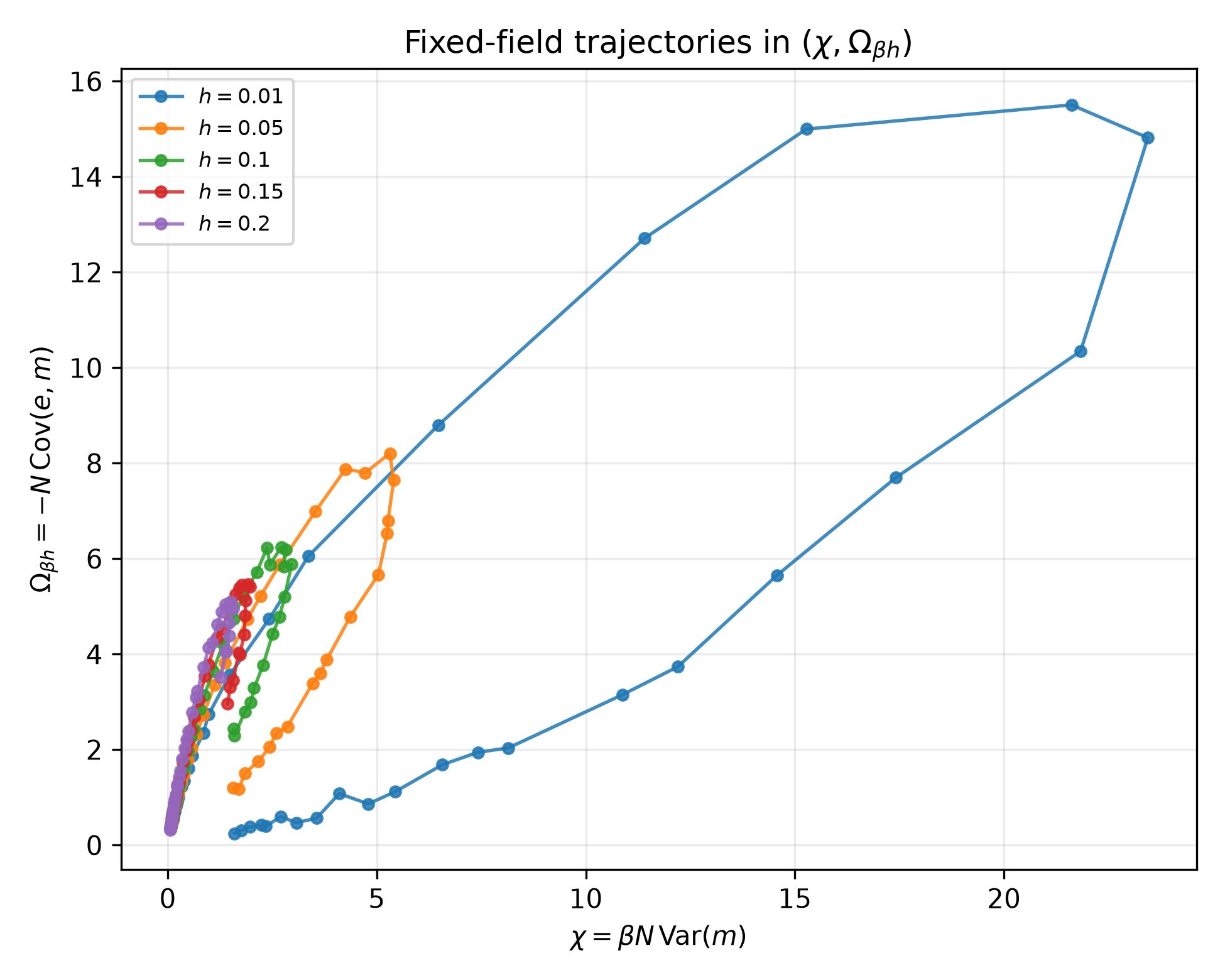}
\caption{
\erbedit{
Thermodynamic trajectories in response space obtained by plotting
$|\Omega_{\beta h}|$ as a function of susceptibility $\chi$
for several magnetic fields. Each curve is parameterized by
inverse temperature $\beta$.
}
}
\label{fig:omega_chi_trajectories}
\end{figure}

\erbedit{
This interpretation provides a natural explanation for the structure of the response-space trajectories. If the ratio $\Omega_{\beta h}/\chi$ governs the local direction of constant-magnetization contours, then the relationship between the susceptibility and mixed response field is not arbitrary. Rather, both quantities are projections of the same underlying response geometry. The trajectories shown in Fig.~\ref{fig:omega_chi_trajectories} may therefore be viewed as integral curves generated by the coupled evolution of magnetic and energetic fluctuations as the critical region is traversed.
}

\erbedit{
To test whether this geometric relationship persists across different magnetic fields, we normalize each response function by its corresponding maximum value,
}
\begin{equation}
\tilde{\chi}
=
\frac{\chi}{\chi_{\max}},
\qquad
\tilde{\Omega}
=
\frac{|\Omega_{\beta h}|}
{|\Omega_{\beta h}|_{\max}}.
\end{equation}

\erbedit{
This normalization removes the trivial field dependence associated with the overall magnitude of the response functions and isolates the intrinsic shape of the thermodynamic trajectory.
}

\erbedit{
Figure~\ref{fig:omega_chi_trajectories} reveals a strong correlation between the magnetic and mixed response sectors. Although the trajectories differ quantitatively for different fields, they exhibit a common overall structure, suggesting that the covariance response is not an independent thermodynamic quantity but is instead constrained by the evolution of the susceptibility.
}

\erbedit{
To separate trivial amplitude differences from geometric structure, we normalize each response function by its peak value,
}
\begin{equation}
\tilde{\chi}
=
\frac{\chi}{\chi_{\max}},
\qquad
\tilde{\Omega}
=
\frac{|\Omega_{\beta h}|}
{|\Omega_{\beta h}|_{\max}}.
\end{equation}

\begin{figure}[t]
\centering
\includegraphics[width=\columnwidth]{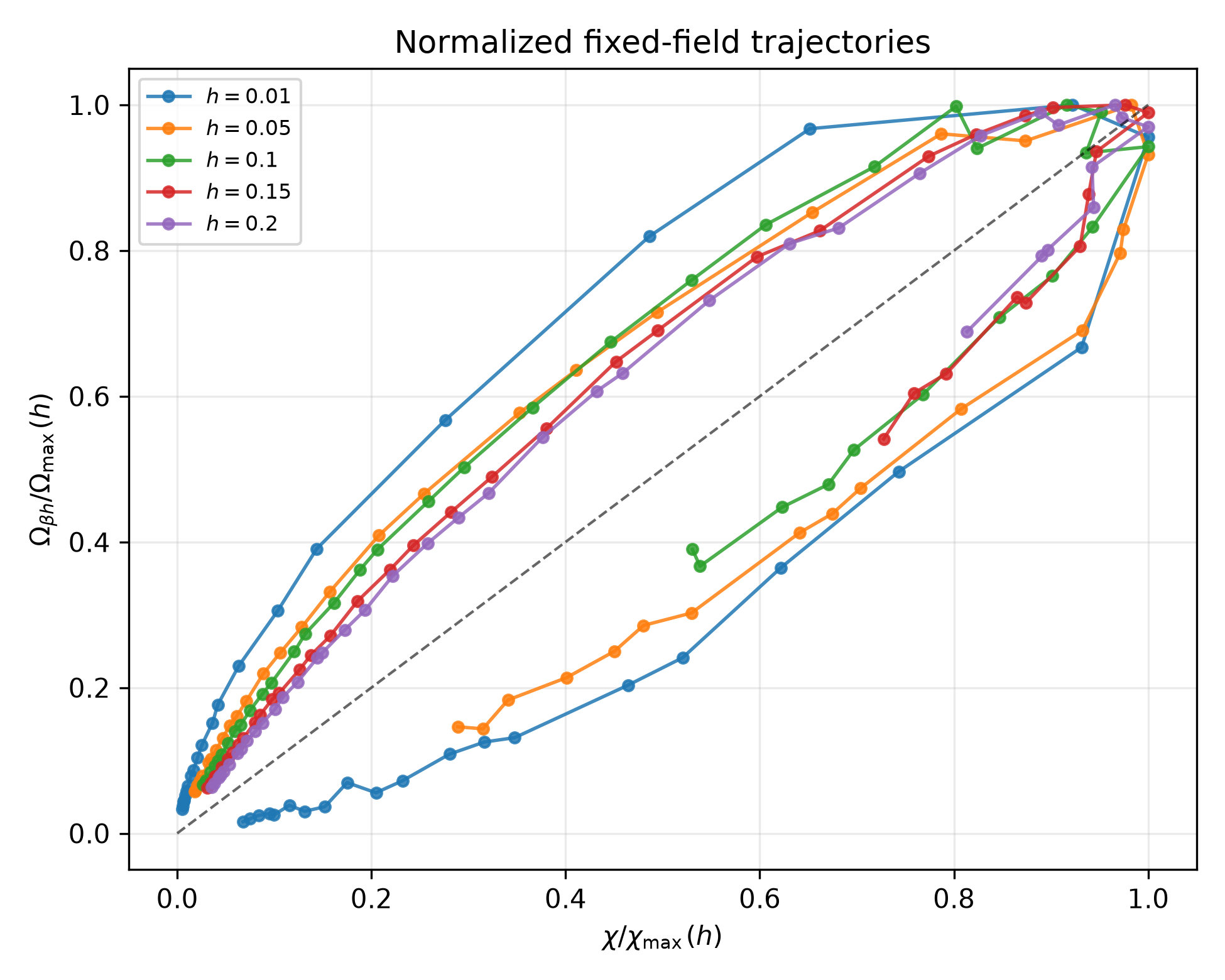 }
\caption{
\erbedit{
Normalized thermodynamic trajectories in response space obtained by plotting
$\tilde{\Omega}=|\Omega_{\beta h}|/|\Omega_{\beta h}|{\max}$
as a function of
$\tilde{\chi}=\chi/\chi{\max}$
for several magnetic fields. Each trajectory is parameterized by inverse temperature $\beta$, with increasing $\beta$ corresponding to evolution from the disordered high-temperature regime toward the ordered phase. Despite substantial differences in the absolute magnitudes of the response functions, the normalized trajectories collapse onto a common curve, indicating that the mixed response and susceptibility are governed by a shared underlying response geometry. The approximate field independence of the normalized trajectories suggests the emergence of a universal thermodynamic manifold in response space near criticality.
}
}
\label{fig:omega_chi_normalized}
\end{figure}

\erbedit{
The resulting collapse is striking. Although the individual response functions differ substantially in magnitude and peak location, the normalized trajectories follow nearly the same path through response space. This behavior indicates that the mixed response field and susceptibility are not independent observables but instead evolve according to a common geometric constraint. In this representation, the critical region is characterized not by a single diverging quantity but by a well-defined trajectory that appears largely insensitive to the magnitude of the external field.
}

\erbedit{
The existence of such a trajectory suggests that the critical crossover may be described by an emergent response manifold whose structure is encoded in the relative evolution of the fluctuation observables. From this perspective, the susceptibility and mixed response field play roles analogous to coordinates on a reduced thermodynamic manifold, while the normalized trajectory describes the path traced out by the system as it approaches criticality.
}

\erbedit{
Viewed collectively, the normalized trajectories occupy a
remarkably restricted region of response space. Rather than
filling a two-dimensional domain, the data collapse onto a
narrow curve-like structure, indicating that the mixed
response field and susceptibility remain strongly constrained
throughout the crossover region. The effective dimensionality
of the response is therefore substantially reduced relative
to that of the underlying control-parameter space.
}

\erbedit{
This observation suggests the existence of an emergent
response manifold on which the dominant thermodynamic
fluctuations evolve. The manifold is not defined by the
equilibrium state variables themselves, but by the relations
among the response functions that characterize fluctuations
about those states. In this sense, the structure revealed
here differs fundamentally from traditional thermodynamic
surfaces such as equations of state or coexistence curves.
Instead, it describes the geometry of response.
}

\erbedit{
The appearance of a low-dimensional response manifold is
consistent with the differential relation
}
\begin{align}
\left(\frac{dh}{d\beta}\right)_m
=
-\frac{\Omega_{\beta h}}{\chi},
\end{align}
\erbedit{
which identifies the ratio of mixed and magnetic response
functions with the local slope of constant-magnetization
contours in control-parameter space. The collapse of the
normalized trajectories therefore implies that these local
directions acquire a nearly universal structure near
criticality. From this perspective, the response manifold
may be viewed as the image of a family of isomagnetization
flows projected into response space.
}

\erbedit{
The collapse of the normalized trajectories suggests that
the response map
$\Phi:(\beta,h)\mapsto(\chi,\Omega_{\beta h})$
possesses an underlying geometric organization.
Although the thermodynamic control manifold is
two-dimensional, its image in normalized response space
becomes concentrated near an approximately one-dimensional
curve. This behavior is consistent with the existence of
an emergent response structure that constrains the joint
evolution of the susceptibility and mixed response field.
Because Eq.~(\ref{eq:isomag}) directly relates the ratio
$\Omega_{\beta h}/\chi$ to the local direction of
constant-magnetization trajectories, the observed collapse
suggests that these response-space relations reflect
underlying geometric constraints on the thermodynamic
control manifold. In this sense, the response ridges and
trajectory relations identified here point toward a
response geometry that is distinct from, but related to,
the geometry of thermodynamic state space.
}


\subsection{Geometric Organization of Response}

\erbedit{
The structures identified above differ fundamentally from
the geometric constructions traditionally employed in
thermodynamics. Classical thermodynamic geometry is defined
on a manifold of equilibrium states, where metric tensors
constructed from thermodynamic potentials characterize
fluctuations, stability, and critical behavior
\cite{Weinhold1975,Ruppeiner1995}. In these approaches the
geometry is a property of the state manifold itself.
}

\erbedit{
The response structures identified here arise from a
different construction. Rather than characterizing the
geometry of states, they characterize the geometry of
response. The primary objects are not state variables such
as entropy, energy, or magnetization, but the derivatives
that describe how these quantities respond to changes in
external control parameters.
}

\erbedit{
For the Ising system, the control manifold is spanned by the
coordinates $(\beta,h)$, while the response manifold is
spanned by the observables
$(\chi,\Omega_{\beta h})$. The response functions therefore
define a map
}
\begin{equation}
\Phi:
(\beta,h)
\longrightarrow
(\chi,\Omega_{\beta h}).
\end{equation}

\erbedit{
The normalized trajectories discussed above may be viewed as
the image of this map projected into response space. The
observed trajectory collapse implies that the image of the
two-dimensional control manifold becomes concentrated near a
lower-dimensional structure, suggesting the existence of an
emergent response manifold governing the dominant
fluctuations.
}

\erbedit{
This viewpoint parallels the geometric framework developed
for nonequilibrium thermodynamic response in open quantum
systems\cite{Bittner2026_jcp} There, response is naturally decomposed into
symmetric and antisymmetric sectors,
}
\begin{equation}
R_{\mu\nu}
=
G_{\mu\nu}
+
F_{\mu\nu},
\end{equation}
\erbedit{
where the symmetric component $G_{\mu\nu}$ defines a metric
structure associated with fluctuations and susceptibility,
while the antisymmetric component $F_{\mu\nu}$ defines a
curvature associated with path-dependent response and work.
Although the present Ising analysis concerns an equilibrium
system, the coexistence of susceptibility and mixed-response
sectors suggests an analogous separation between fluctuation
geometry and response geometry.
}

\erbedit{
From this perspective, criticality may be viewed as the
emergence of geometric structure in response space. The
traditional state-based description emphasizes singularities
of thermodynamic functions and their derivatives. The
response-based description instead emphasizes the relations
among response functions themselves. The collapse of the
normalized trajectories indicates that these relations
acquire a remarkably simple form near criticality, revealing
an emergent geometry that is largely independent of the
magnitude of the applied field.
}

\erbedit{
The emergence of a reduced response manifold is particularly
interesting in light of recent efforts to formulate
thermodynamic response in geometric terms. While the present
analysis is confined to the equilibrium Ising model, the
observed trajectory collapse suggests that relations among
response functions may provide an alternative route to
identifying geometric structure in many-body systems. Whether
similar response manifolds arise in more general equilibrium
and nonequilibrium settings remains an important question for
future work.
}
\section{Discussion}\label{sec:Discussion}

The results presented here reveal a geometric organization of thermodynamic response
that extends beyond conventional analyses based on individual response functions. While
the susceptibility and heat capacity characterize magnetic and energetic fluctuations
separately, the mixed response field $\Omega_{\beta h}$ probes the coupling between
these fluctuation sectors. As a consequence, it provides information that is not
contained in either response function alone and offers a complementary perspective on
critical behavior.

The mixed response develops a pronounced ridge structure that emanates from the
critical point and extends into the finite-field crossover regime. Similar crossover
trajectories have long been associated with Widom lines and response maxima in fluids
and magnetic systems. From the present perspective, however, these structures arise
naturally as geometric features of the response landscape. The maxima of $\chi$,
$C_h$, and $\Omega_{\beta h}$ define distinct response ridges whose locations and
scaling behavior encode how different fluctuation sectors evolve away from criticality.
The observation that these ridges possess different scaling exponents demonstrates that
magnetic, energetic, and mixed fluctuations organize differently throughout the
crossover region.

Beyond the ridge structure itself, the normalized response trajectories reveal an
additional level of organization. When expressed in terms of scaled response variables,
data obtained over a broad range of magnetic fields collapse onto an approximately
universal curve. This collapse indicates that the response functions are not independent
but are constrained by underlying relations that persist throughout the crossover
regime. In geometric terms, the trajectories occupy a low-dimensional region of
response space despite originating from a two-dimensional control manifold. The
emergence of such structure suggests that critical fluctuations evolve on an effective
response manifold whose geometry is determined by the interplay of thermal, magnetic,
and mixed response.

The trajectory relations derived in Sec.~\ref{sec:IsingModel} provide a possible
interpretation of this behavior. Both isentropic and constant-magnetization trajectories
are governed by ratios of response functions, such as $\Omega_{\beta h}/\chi$ and
$\Omega_{\beta h}/C_h$. The latter ratio is recognized as the magnetic
Gr\"{u}neisen parameter $\Gamma_h \equiv \Omega_{\beta h}/C_h$, which characterizes
the magnetocaloric response and is known to diverge universally at critical points with
an exponent set by the universality
class.\cite{ZhuGarstRoschSi2003,GarstRosch2005,WuZhuSi2018} Near criticality,
$\Omega_{\beta h}$, $\chi$, and $C_h$ are all governed by the same scaling variable
$h/|t|^{\Delta}$, so that their ratios $\Gamma_h$ and $\Omega_{\beta h}/\chi$ acquire
a nearly universal form throughout the crossover region. The observed collapse of
normalized response trajectories in Fig.~5 is a direct consequence of this constraint:
because both $\chi$ and $\Omega_{\beta h}$ are scaling functions of the same argument,
their normalized ratio traces a common curve independent of field strength. The
response ridges and trajectory collapse may therefore be viewed as manifestations of a
common geometric structure generated by the response functions themselves, with the
Gr\"{u}neisen parameter providing the connecting link between magnetocaloric response,
critical scaling, and the geometry of the crossover region.

More broadly, the present results suggest a distinction between the geometry of
thermodynamic states and the geometry of thermodynamic response. Traditional geometric
approaches, including information geometry and fluctuation metrics, characterize
properties of equilibrium states. The framework developed here instead focuses on
relationships among response functions defined over a control manifold. In this view,
geometry emerges through the organization of fluctuations, the trajectories they
induce, and the constraints that relate different response sectors.

A related connection between geometry and crossover phenomena has been explored
previously using information-geometric methods, where geometric invariants constructed
from equilibrium thermodynamic metrics have been shown to track Widom lines and
associated crossover structures in both magnetic and fluid
systems.\cite{DeyRoySarkar2013} In such approaches, the geometry is derived from the
properties of equilibrium states and their fluctuation metrics. The present framework
adopts a different viewpoint. Rather than characterizing the geometry of thermodynamic
states, it focuses on the organization of response functions defined over a control
manifold. The response ridges, trajectory relations, and scaling structures identified
here therefore arise from the geometry of thermodynamic response itself rather than
from a metric defined on equilibrium state space.

It is instructive to compare the present framework with the Ruppeiner geometric
approach, in which a metric tensor is defined on the manifold of equilibrium states by
taking the Hessian of the entropy with respect to extensive
variables.\cite{Ruppeiner1995} The resulting scalar curvature $R$ is a nonlinear,
second-order combination of the thermodynamic response functions $C_h$, $\chi$, and
$\Omega_{Th}$, together with their derivatives, and diverges at the critical point with
an exponent proportional to the correlation volume $\xi^d$.\cite{Ruppeiner1995,Ruppeiner2010}
The locus of $|R|$ maxima in the $(h,T)$ plane has been used to define Widom lines in
magnetic and fluid systems.\cite{DeyRoySarkar2013,RuppeinerSahaySarkarSengupta2012}

The mixed response field $\Omega_{\beta h}$ is related to but distinct from $R$.
Although $\Omega_{\beta h} = -(\partial M/\partial T)_h$ appears as an off-diagonal
element of the Ruppeiner metric in appropriate thermodynamic coordinates, it is a
\emph{first-order} object — a single mixed derivative of the free energy — whereas $R$
is constructed from the full metric and its second derivatives. More fundamentally, the
two approaches differ in the geometric space they consider. The Ruppeiner construction
defines geometry on the manifold of \emph{equilibrium states}, where the coordinates
are thermodynamic state variables such as energy and magnetization. The present
framework instead defines geometry on the \emph{control manifold} $(\beta, h)$ of
externally controllable parameters, where $\Omega_{\beta h}$ arises as the curvature
of the response one-form $\mathcal{A} = -M\,dh$. This distinction is not merely
technical: in the Ruppeiner picture, geometry is a property of equilibrium states; in
the present picture, geometry emerges from the organization of response functions and
their mutual relationships under changes in external control parameters. As a
consequence, the locus of $|\Omega_{\beta h}|$ maxima — the mixed-response ridge
studied here — is generically distinct from the locus of $|R|$ maxima, and the two
constructions provide complementary rather than redundant information about the
crossover structure near the critical point.

The central result of this work is not the numerical value of a scaling exponent, nor
the identification of a particular crossover ridge. Rather, it is the recognition that
the mixed response field
\begin{equation}
\Omega_{\beta h} = -N\,\mathrm{cov}(m,e)
\end{equation}
possesses a direct geometric interpretation. While the susceptibility and heat capacity
characterize fluctuations within the magnetic and energetic sectors separately, the
mixed response quantifies how those sectors are coupled. The covariance between energy
and magnetization is therefore not merely a statistical quantity; it encodes how thermal
and magnetic fluctuations jointly organize the thermodynamic response landscape.

This interpretation becomes apparent through the geometric structures associated with
$\Omega_{\beta h}$. The mixed response generates its own ridge structure, exhibits
scaling behavior distinct from that of either $\chi$ or $C_h$, and governs the
direction of thermodynamic trajectories through relations such as
\begin{equation}
\left(\frac{dh}{d\beta}\right)_m = -\frac{\Omega_{\beta h}}{\chi}.
\end{equation}
The resulting trajectory collapse demonstrates that correlated fluctuations constrain
the evolution of the system in a manner that is not visible from either response
function alone. From this perspective, the covariance field acts as a geometric
mediator between magnetic and energetic response, revealing an organization of critical
fluctuations that would otherwise remain hidden.

The broader implication is that mixed fluctuations may carry geometric information that
is absent from conventional response functions. In the present case, the covariance
between energy and magnetization induces a structured response landscape characterized
by ridges, crossover trajectories, and low-dimensional response organization. The
geometry therefore emerges not from the equilibrium states themselves, but from the
correlations that couple distinct fluctuation sectors. This observation suggests a more
general principle: geometric structure in thermodynamic systems may be encoded not only
in fluctuations themselves, but in the covariances that relate them.

Several directions for future work naturally follow from this perspective. An immediate
extension is to investigate whether similar response-manifold structures arise in other
magnetic models, fluids, and systems exhibiting continuous phase transitions. It will
also be interesting to determine whether the observed trajectory collapse can be derived
from scaling theory or renormalization-group arguments. More generally, the present
framework suggests the possibility of constructing geometric descriptions directly in
response space, where response ridges, fluctuation couplings, and trajectory relations
provide the fundamental objects of interest.

Taken together, these results indicate that mixed thermodynamic response provides more
than an additional response function. It reveals a geometric organization of critical
fluctuations that links response ridges, scaling behavior, and crossover trajectories
within a common framework. The resulting response geometry complements traditional
state-space descriptions and provides a new perspective on the structure of critical
phenomena.

\vspace{0.5cm}

\section*{Data Availability Statement}
All data generated or analyzed during this study are included in this manuscript.

\begin{acknowledgments}
ERB acknowledges funding from the National Science Foundation (CHE-2404788), Robert A.\ Welch Foundation (E-1337), the Department of Energy supported this research through Award No. 11937-PO147716. ERB gratefully acknowledges funding from IVADO for a Visiting Professorship at the Institut Courtois, Universit\'e de Montr\'eal. 
\end{acknowledgments}

\section*{Author Contributions}
The author developed the theoretical framework, performed all derivations,
and carried out the analysis presented in this work. Generative AI tools
were used during manuscript preparation to assist with code development
 and verification of intermediate algebraic
steps. All theoretical formulations, results, and physical
interpretations were independently developed and validated by the author.

\section*{Conflicts of Interest}
The author declares no competing financial or non-financial interests.

\bibliographystyle{apsrev4-2}
\bibliography{refs_consolidated}

\newpage


\end{document}